\documentclass[aps, twocolumn, prl, superscriptaddress]{revtex4-2}
\usepackage{amsmath}
\usepackage{bm}
\usepackage{mathtools}
\usepackage{physics}
\usepackage{datetime}
\usepackage{comment}
\newdate{date}{31}{05}{2023}
\usepackage{enumerate}
\usepackage{graphicx} 
\usepackage[hidelinks]{hyperref} 
\usepackage{bibunits} 
\defaultbibliographystyle{apsrev4-2}
\defaultbibliography{bibliography}

\makeatletter
\def\maketitle{
\@author@finish
\title@column\titleblock@produce
\suppressfloats[t]}
\makeatother

\begin{document}

\title{Time-resolved pairing gap spectroscopy in a quantum simulator of fermionic superfluidity inside an optical cavity}

\author{Dylan J. Young}
\affiliation{JILA, NIST, and Department of Physics, University of Colorado, Boulder, CO, USA}
\author{Eric Yilun Song}
\affiliation{JILA, NIST, and Department of Physics, University of Colorado, Boulder, CO, USA}
\author{Anjun Chu}
\affiliation{JILA, NIST, and Department of Physics, University of Colorado, Boulder, CO, USA}
\affiliation{Center for Theory of Quantum Matter, University of Colorado, Boulder, CO, USA}
\author{Diego Barberena}
\affiliation{JILA, NIST, and Department of Physics, University of Colorado, Boulder, CO, USA}
\affiliation{Center for Theory of Quantum Matter, University of Colorado, Boulder, CO, USA}
\author{Zhijing Niu}
\affiliation{JILA, NIST, and Department of Physics, University of Colorado, Boulder, CO, USA}
\author{Vera M. Sch\"afer}
\affiliation{JILA, NIST, and Department of Physics, University of Colorado, Boulder, CO, USA}
\affiliation{Max-Planck-Institut f\"ur Kernphysik, Saupfercheckweg 1, 69117 Heidelberg, Germany}
\author{Robert J. Lewis-Swan}
\affiliation{Homer L. Dodge Department of Physics and Astronomy, University of Oklahoma, Norman, OK, USA}
\affiliation{Center for Quantum Research and Technology, University of Oklahoma, Norman, OK, USA}
\author{Ana Maria Rey}
\affiliation{JILA, NIST, and Department of Physics, University of Colorado, Boulder, CO, USA}
\affiliation{Center for Theory of Quantum Matter, University of Colorado, Boulder, CO, USA}
\author{James K. Thompson}
\affiliation{JILA, NIST, and Department of Physics, University of Colorado, Boulder, CO, USA}
\usdate
\date{\today}

\begin{abstract}
We use an ensemble of laser-cooled strontium atoms in a high-finesse cavity to cleanly emulate the technique of rf spectroscopy employed in studies of BEC-BCS physics in fermionic superfluids of degenerate cold gases.
Here, we leverage the multilevel internal structure of the atoms to study the physics of Cooper pair breaking in this system.
In doing so, we observe and distinguish the properties of two distinct many-body gaps, the BCS pairing gap and the spectral gap, using nondestructive readout techniques.
The latter is found to depend on the populations of the internal atomic states, reflecting the chemical potential dependence predicted in fermionic superfluids.
This work opens the path for more fully exploiting the rich internal structure of atoms in cavity QED emulators to study both analogous systems and also more exotic states yet to be realized.
\end{abstract}

\maketitle
\vskip 0.5cm

\begin{bibunit}
Ultracold atomic systems have emerged as a promising platform for simulating theories of quantum many-body physics, benefiting from the capability to engineer clean and tunable interactions of many forms \cite{Gross2017,browaeys_2020_natphys,altman_2021_prxquantum,mivehvar_2021_advphys,chomaz_2022_repprogphys,defenu_2023_revmodphys}. An additional advantage lies in the complex internal structure of cold atoms, which have multiple ground and excited states and can lie beyond the paradigm of a traditional qubit.
Such multilevel systems open up an even broader class of models and physical systems for study \cite{chomaz_2022_repprogphys,stamper-kurn_2013_rmp,zhang_2014_science,hebenstreit_2017_prl,asenjo-garcia_2019_pnas,pineiro-orioli_2019_prl_rey,hemmer_2021_pra,chu_2023_prr_amr,orioli_2022_prx_amr,sundar_2023_prl_rey,sundar_2024_prl_amr,cooper_2024_natphys}.

One iconic example is the development of ultracold Fermi gases using neutral atoms.
Such experiments have enabled groundbreaking studies of the phenomenon of fermionic superfluidity, including a first realization of the BCS-BEC crossover \cite{demarco_1999_science_jin,regal_2004_prl_jin,zwierlein_2004_prl,Randeria:2013kda},
and have provided insight into a broad range of many-body systems \cite{chamel_2017_jastrophysastron,chen_2024_revmodphys}.
To probe the superfluid pairing gap, early experiments relied on a ``radio-frequency (rf) spectroscopy'' technique,
which involved weakly driving a mixture of two spin states along an rf transition to a (nominally non-interacting) third state \cite{Regal2003,Greiner2005,gupta2003radio}.
This additional degree of freedom allowed researchers to observe an energy shift related to breaking Cooper pairs \cite{Torma2000}.
While this technique has been widely used, undesired effects, such as competing pair-breaking processes and unwanted interactions between the third state and the other two states,
complicated efforts to analyze the pairing gap \cite{He2005,Punk2007,Giorgini2008}.
Additionally, the predicted shift, sometimes called the spectral gap $\Delta_\mathrm{SG}$, \cite{shin_2007_prl}, depends nontrivially on the chemical potentials of the component spin states and is not equivalent to the pairing gap $\Delta_\mathrm{BCS}$ \cite{Torma2000, Randeria:2013kda}.

Inspired by this context, here we extend earlier work studying BCS superconductivity in a cavity QED platform \cite{lewis-swan_2021_prl_amr,young_2024_nature_jkt} by leveraging multiple internal levels of $^{88}$Sr atoms. We start with implementing the so-called Anderson pseudospin mapping \cite{Anderson1958}, mapping the presence and absence of a Cooper pair onto a long-lived electronic transition. The atoms, which are collectively coupled to a detuned optical cavity, experience an effective all-to-all interaction \cite{norcia_2018_science_jkt} which emulates the attractive and collective interaction between electrons in different momentum states in a superconductor. With this setup, we have previously studied the BCS Hamiltonian and observed dynamical phases in the pairing gap $\Delta_\mathrm{BCS}$ \cite{young_2024_nature_jkt}. By coupling to a third atomic state, we can mimic the setup of rf spectroscopy experiments and measure $\Delta_\mathrm{SG}$. Beyond the analogy to superfluidity, this system is also closely related to donor-acceptor models of energy transfer in biological systems, where the internal atomic states map onto donor and acceptor sites \cite{dong_2012_lsa_sun,gorman_2018_prx_haeffner,potocnik_2018_natcomm_wallraff}.

We construct an appropriate three-level system using an applied magnetic field to couple excited Zeeman states. By tuning the magnetic field strength, we explore this coupling both in a gapped regime and in a strong-coupling regime featuring large population transfer between states. Moreover, by varying the atomic inversion along the initial two-level system, we change both the number of particles participating in the pairing as well the pairing strength, allowing us to explore the distinction between $\Delta_\mathrm{SG}$ and $\Delta_\mathrm{BCS}$. We accomplish all this through the use of two real-time probes, including a novel nondestructive, large dynamic range measurement of the cavity resonance frequency.

\begin{figure}[t]
    \includegraphics[keepaspectratio, width=\columnwidth]{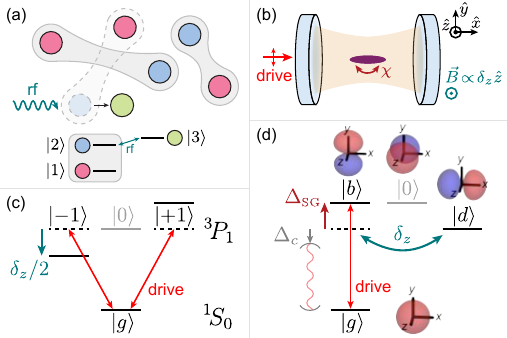}
    \caption{Experimental setup.
    (a) The technique of ``rf spectroscopy'' used in ultracold fermion experiments measures a frequency shift, associated with breaking a Cooper pair, along an auxiliary transition.
    (b) In our system, $^{88}$Sr atoms experience an effective infinite-range spin-exchange interaction through detuned collective coupling to an optical cavity, oriented along $\hat{x}$. We apply a tunable magnetic field along $\hat{z}$ and address the atoms with a $\hat{y}$-polarized laser drive.
    (c) We explore a three-level system consisting of $\ket{g} =\ket{\!\,^{1}\! S_{0}}$ and $\ket{\pm 1} = \ket{^{3}P_{1}, m_j=\pm 1}$. We define $\delta_z$ as the splitting between $\ket{\pm 1}$ states.
    (d) The drive and cavity couple to the effective two-level system $\ket{g}\leftrightarrow\ket{b}$. In this basis, $\delta_z$ couples $\ket{b}$ to a third state $\ket{d}$, analogous to the level structure used in rf spectroscopy experiments. The cavity is detuned from the atomic transition by $\Delta_c = \omega_{c0} - \omega_a$; this generates a collective shift $\Delta_\mathrm{SG}$ of $\ket{b}$, representing the spectral gap. Toy models of the atomic orbitals are shown to highlight the different polarizations required to address the excited states.
    } \label{fig1}
\end{figure}

For these experiments, we cool and trap $N=10^5 - 10^6$ $^{88}$Sr atoms into a high-finesse optical cavity in the strong collective coupling regime, described in previous work \cite{norcia_2016_sciadv_jkt,muniz2020exploring}. The cavity's unshifted resonance frequency $\omega_{c0}$ is red-detuned from the bare $^{3}P_{1}-\,\!\!^{1}S_0$ transition frequency $\omega_a$ by $\Delta_c/2\pi \coloneqq (\omega_{c0}-\omega_a)/2\pi = -51$ MHz.
The relevant states include a single ground state $\ket{g} \coloneqq \ket{^{1}S_{0}}$ and two excited Zeeman states $\ket{\pm 1} \coloneqq \ket{^{3}P_{1}, m_j=\pm 1}$, with a quantization axis along $\hat{z}$ using the coordinates in Fig.~\hyperref[fig1]{\ref{fig1}(b)}. A $\hat{y}$-polarized laser drive excites the ``bright'' superposition $\ket{b}_{k} \coloneqq \tfrac{1}{\sqrt{2}} \left( \ket{+1}_{k} + \ket{-1}_{k} \right)$ for each atom $k$, which also couples to the $\hat{y}$-polarized cavity mode. The orthogonal ``dark'' state $\ket{d}_{k} \coloneqq \tfrac{1}{\sqrt{2}} \left( \ket{+1}_{k} - \ket{-1}_{k} \right)$ has a spatial distribution aligned with the cavity axis ($\hat{x}$), as shown in Fig.~\hyperref[fig1]{\ref{fig1}(d)}, and thus does not radiate into the cavity.
In this natural basis for our system, we define collective dipole operators $\hat{J}^{+} = (\hat{J}^{-})^\dagger \coloneqq \sum_{k=1}^N \ketbra{b}{g}_{k}$, measuring dipole projections along $\hat{y}$, and the atomic inversion $\hat{J}^{z} = \sum_{k=1}^N \hat{J}_{k}^{z}$ where $\hat{J}_{k}^{z} \coloneqq \tfrac{1}{2} (\ketbra{b}{b}_{k} -\ketbra{g}{g}_{k})$.
We also apply a tunable, uniform magnetic field $\vec{B}\parallel\hat{z}$, represented using the collective angular momentum operator $\hat{L}^{z} \coloneqq \sum_{k=1}^{N} \tfrac{1}{2}(\ketbra{+1}{+1}_{k} - \ketbra{-1}{-1}_{k}) = \sum_{k=1}^{N} \tfrac{1}{2}(\ketbra{b}{d}_{k} + \ketbra{d}{b}_{k})$ which describes linear Zeeman shifts.
 
Using the above definitions, up to atom-light coupling inhomogeneities neglected here for simplicity \cite{supp}, we engineer an effective atomic Hamiltonian of the form:

\begin{equation}
    \hat{H}_{\mathrm{a}}/\hbar = -\chi \hat{J}^{+}\hat{J}^{-} + \delta_z \hat{L}^{z} +\sum_{k=1}^N \varepsilon_{k} \ketbra{g}{g}_{k}.
    \label{eq:ham_a}
\end{equation}

\noindent The first term describes an all-to-all cavity-mediated spin-exchange interaction with strength $\chi = g^2/\Delta_c$, where $2g/2\pi = 15.4$~kHz is the rms atom-cavity coupling as a single-photon Rabi frequency. Since this interaction is collective, it scales with atom number with characteristic strength $\chi N$. $\delta_z$ is the tunable splitting between $\ket{\pm 1}$ Zeeman states; in the bright/dark basis, it creates an effective torque that rotates the collective dipole moment and couples $\ket{b}$ and $\ket{d}$. Finally, $\varepsilon_{k}$ describes the differential light shift between the ground state $\ket{g}$ and the excited states due to the intracavity trapping light at $813$~nm for atom $k$. This shift varies between the atoms due to a finite motional distribution \cite{young_2024_nature_jkt}, which induces dephasing between the ground and excited states.

The above Hamiltonian has a direct analogy to rf spectroscopy experiments using ultracold fermions in the BCS limit. Following an Anderson pseudospin mapping \cite{Anderson1958}, we map the $\ket{b}$ and $\ket{g}$ states of atom $k$ onto the presence and absence of a Cooper pair of fermions with different internal states ($\ket{1}$ and $\ket{2}$ in Fig.~\hyperref[fig1]{\ref{fig1}(a)}) and momenta $\pm\mathbf{k}$. The spin-exchange term in Eq.~\ref{eq:ham_a} then represents the BCS interaction between fermions, and the $\varepsilon_k$ term describes the electrons' kinetic energy. Additionally, the (complex) BCS pairing gap $\Delta_{\mathrm{BCS}}$ corresponds to $\chi\langle\hat{J}^{-}\rangle$. We can go further by mapping an atom in state $\ket{d}$ onto a ``broken'' Cooper pair consisting of one fermion in $\ket{1}$ and the other transferred to a third state $\ket{3}$. The Zeeman term coupling $\ket{b}$ to $\ket{d}$ then represents a drive which breaks up Cooper pairs \cite{supp}. Unlike the drive in rf spectroscopy experiments, this coupling operates at DC; we tune the coupling strength, rather than an rf drive frequency, in order to probe features of the spectral gap $\Delta_\mathrm{SG}$.

To study this model, we initialize the atoms with a collective drive angle $\theta_0 = \pi/2$ (defined as a $\pi/2$ pulse for maximally-coupled atoms) along the $\ket{g}$ to $\ket{b}$ transition using a $<250$~ns laser drive pulse. At time $t=0$, we turn off the drive. As the atoms evolve under Eq.~\ref{eq:ham_a}, they weakly emit $\hat{y}$-polarized light into the far-detuned cavity which adiabatically follows $\langle\hat{J}^{-}\rangle$. As outlined in earlier work \cite{young_2024_nature_jkt}, this allows us to measure $\langle\hat{J}^{-}\rangle$ in real time by measuring this light as it leaks out of the cavity using a heterodyne detector, with only a small fraction of atoms emitting light (thus avoiding a large backaction on the ensemble).


\begin{figure}[t]
    \includegraphics[keepaspectratio, width=\columnwidth]{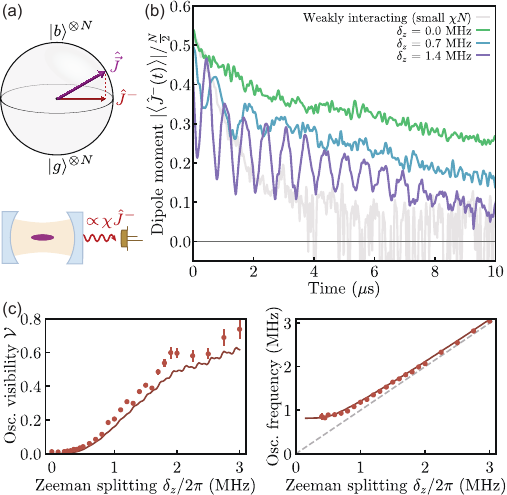}
    \caption{Oscillations in the pairing gap $\Delta_{\mathrm{BCS}}$ from excited state coupling.
    (a) The atoms emit light weakly into the cavity at a rate proportional to $\Delta_{\mathrm{BCS}}=\chi \langle \hat{J}^{-} \rangle$, which we detect to infer dynamics.
    (b) Time traces of $\lvert\langle \hat{J}^{-} \rangle\rvert$ with $\chi N/2\pi$ = 1~MHz. The Zeeman splitting $\delta_z$ induces rotations in the collective dipole moment, which are suppressed for small $\delta_z$. The gray trace represents single-particle dynamics (with $\delta_z/2\pi = 0$~MHz), using a smaller $N$ to weaken the collective interactions.
    (c) The average oscillation visibility (left) and frequency (right) from $t=0.5~\mu$s to $t=5~\mu$s for different $\delta_z$. We also plot numerical simulations (solid curves) and the single-particle response $\omega_\mathrm{osc} = \delta_z$ (gray dashed line).
    All error bars in the paper represent $\pm 1\sigma$ deviations over a bootstrap resampling on experimental shots ($n_\mathrm{boot}$ = 100).} \label{fig2}
\end{figure}

We first explore the competition between spin-exchange interactions and excited state coupling in Fig.~\hyperref[fig2]{\ref{fig2}(b)} by comparing the time dynamics of the BCS pairing gap $\lvert\Delta_\mathrm{BCS}\rvert\propto \lvert\langle \hat{J}^{-} \rangle\rvert$ for different coupling strengths $\delta_z$. Here, we scan $\delta_z/2\pi$ from $0$ MHz to $5$ MHz between shots while holding a fixed characteristic interaction strength $\chi N/2\pi = 1.0$~MHz. A background inhomogeneity set by $\{ \varepsilon_{k} \}$ with standard deviation $\varepsilon/2\pi = 150$~kHz also remains fixed; in the absence of interactions, this sets a dephasing time of $1~\mu$s as shown by the gray trace. Consistent with previous results \cite{young_2024_nature_jkt}, we observe that sufficiently large interactions protect against dephasing at a scale set by $\Delta_\mathrm{BCS}$, drastically enhancing the coherence time. The coupling induced by $\delta_z$ rotates the collective dipole moment, inducing oscillations in $\lvert\langle \hat{J}^{-} \rangle\rvert$. For small $\delta_z$, weak oscillations open up at short times but damp after a few microseconds. In contrast, large $\delta_z$ traces feature large-amplitude oscillations with long lifetimes. The fact that $\lvert\langle \hat{J}^{-} \rangle\rvert$ remains large in this limit at long times shows that the interactions continue to protect against dephasing from the single-particle inhomogeneity $\{\varepsilon_{k}\}$, even with a large $\delta_z$.

Fig.~\hyperref[fig2]{\ref{fig2}(c)} further studies these two regimes by analyzing the oscillations in an interval from $t = 0.5~\mu$s to $t = 5~\mu$s.
When $\delta_z/2\pi \gtrsim 2$~MHz, the oscillation visibility (defined as $\mathcal{V}=\frac{\lvert\langle \hat{J}^{-} \rangle\rvert_\mathrm{max} - \lvert\langle \hat{J}^{-} \rangle\rvert_\mathrm{min}}{\lvert\langle \hat{J}^{-} \rangle\rvert_\mathrm{max} +\lvert\langle\hat{J}^{-} \rangle\rvert_\mathrm{min}}$) is consistently large, and the oscillation frequency $\omega_\mathrm{osc}$ approaches $\delta_z$.
For small $\delta_z$, a different picture emerges: the visibility is suppressed and approaches 0, and $\omega_\mathrm{osc}$ plateaus to a constant frequency.
The experimental data agrees with numerical simulations (solid curves) except for a small absolute scale factor in the oscillation visibility. This is likely related to additional dephasing mechanisms in our system, unaccounted for in numerical simulations, which reduce the measured $\lvert\langle \hat{J}^{-} \rangle\vert$.
The oscillation frequency at small $\delta_z$ can be directly connected to the spectral gap $\Delta_\mathrm{SG}$ observed in degenerate Fermi gas experiments when coupling to a third state.
This is because our measurement of oscillations induced by a quenched DC coupling provides information equivalent to mapping out an rf spectroscopic peak.

\begin{figure}[b]
    \includegraphics[keepaspectratio, width=\columnwidth]{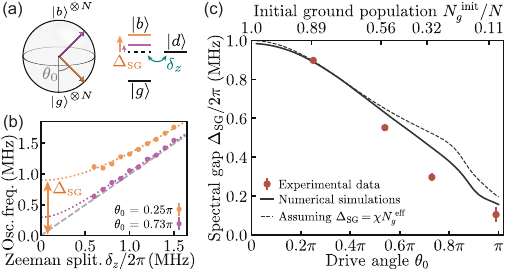}
    \caption{Probing the spectral gap $\Delta_\mathrm{SG}$.
    (a) As the drive angle $\theta_0$ increases, $N_g$ decreases, which is expected to reduce the spectral gap $\Delta_\mathrm{SG}$.
    (b) We repeat the experiment in Fig.~\hyperref[fig2]{\ref{fig2}(c)} for different drive angles $\theta_0$. For each $\theta_0$, we fit the results to the form $\omega_\mathrm{osc} = \sqrt{\Delta_\mathrm{SG}^2 + \delta_z^2}$ (dotted lines) and extract $\Delta_\mathrm{SG}$.
    (c) Plot of $\Delta_\mathrm{SG}$ for different $\theta_0$, or alternatively the $t=0$ ground state atom population $N_{g}^{\mathrm{init}}$ (circles). We compare this against numerical simulations of $\Delta_\mathrm{SG}$ (solid curve) and $\chi N_{g}^{\mathrm{eff}}$ (dashed curve), where $N_{g}^{\mathrm{eff}}$ is the ground state population weighted by cavity coupling and averaged over the measured time interval $t=0.5~\mu$s to $t=5~\mu$s.
    } \label{fig3}
\end{figure}

An intuitive picture of the oscillations can be gained by imagining the atoms as electric dipoles. Atoms with coherence along the $\ket{g}$ to $\ket{b}$ transition exhibit a dipole moment proportional to $\lvert\langle \hat{J}^{-} \rangle\rvert$ which oscillates along $\hat{y}$ at optical frequencies. Applying a magnetic field along $\hat{z}$ creates an effective torque on the dipoles, causing them to rotate in the $xy$\nobreakdash-plane at a frequency set by the Zeeman splitting $\delta_z$. Therefore, the population in $\ket{b}$ should periodically transfer between $\ket{b}$ (aligned with $\hat{y}$) and $\ket{d}$ (aligned with $\hat{x}$) while maintaining coherence with the ground state $\ket{g}$. This effectively causes the collective dipole moment $\lvert\langle \hat{J}^{-} \rangle\rvert$ along $\hat{y}$ to oscillate in time.

However, sufficiently strong spin-exchange interactions of the form $\hat{J}^{+} \hat{J}^{-}$ disrupt this process. This term essentially shifts the excited state $\ket{b}$ by a frequency $\Delta_\mathrm{SG}$ relative to $\ket{d}$, which does not radiate into the cavity.
Quantitatively, $\chi \hat{J}^{+} \hat{J}^{-} = \chi (\hat{J}^2 - (\hat{J}^{z})^2 + \hat{J}^{z})$ has eigenstates $\ket{J,J^{z}}$ and eigenvalues $E_{J,J^{z}}=\chi[J(J+1)-J^{z}(J^{z}-1)]$.
For fully collective states, i.e., where $J = \tfrac{N_b+N_g}{2}$ and $J^{z} = \tfrac{N_b-N_g}{2}$ with $N_{b,g}$ denoting the populations in $\ket{b}$ and $\ket{g}$ respectively, $E_{J,J^{z}} \approx (\chi N_g) N_b$ in the large-$N$ limit.
Therefore, transferring one atom from $\ket{b}$ to $\ket{d}$ imposes an energy penalty equal to $\hbar\Delta_\mathrm{SG} = \hbar\chi N_g$.
Note that $\Delta_\mathrm{SG}$ changes with atomic inversion, mirroring the chemical potential dependence in degenerate Fermi gases.
In contrast, $\Delta_\mathrm{BCS}$, which protects against dephasing induced by $\varepsilon_k$ (the kinetic energy in the BCS system), only depends on the total Bloch vector length $J$ since dephasing preserves $J^{z}$ but changes $J$.

To explore this dependence in our system, we repeat the above experiment with different initial drive angles $\theta_0$ along the $\ket{g}$ to $\ket{b}$ transition. As Fig.~\hyperref[fig3]{\ref{fig3}(a)} illustrates, increasing $\theta_0$ decreases the initialized ground state population $N_{g}^{\mathrm{init}}$, which should reduce the splitting between the excited states and therefore $\Delta_\mathrm{SG}$. We see this trend in Fig.~\hyperref[fig3]{\ref{fig3}(b)}: when $\theta_0 = 0.73\pi$, the oscillation frequency resembles the expected single-particle response, whereas the $\theta_0 = 0.25\pi$ data exhibits large frequency deviations which indicate a sizable spectral gap $\Delta_\mathrm{SG}$.
We quantify $\Delta_\mathrm{SG}$ by fitting the data to the form $\omega_\mathrm{osc} = \sqrt{\Delta_\mathrm{SG}^2 + \delta_z^2}$, the predicted response for homogeneous coupling \cite{supp}, over the domain $\delta_z \geq 0.6~$MHz where the oscillations are most prominent. These fits, shown by the dotted lines, show a large reduction in $\Delta_\mathrm{SG}$ as we increase $\theta_0$.

We verify the relation between $\Delta_\mathrm{SG}$ and the ground state population by calculating $\chi N_g$ from numerical simulations for different $\theta_0$. 
Complications arise in this calculation from inhomogeneous atom-light coupling, since the effective $N_g$ seen by the cavity varies in time (unlike in homogeneously coupled systems where it does not vary on timescales faster than the decay rate).
To account for this, we weight atoms by cavity coupling and average the result over the measurement interval ($t=0.5~\mu$s to $t=5~\mu$s) to obtain an effective shift $\chi N_{g}^{\mathrm{eff}}$ (see Supplemental Material) \cite{supp}, represented by the black dashed curve in Fig.~\hyperref[fig3]{\ref{fig3}(c)}.
By comparing this curve against measurements of $\Delta_\mathrm{SG}$ from experimental data (circles), we demonstrate that $\Delta_\mathrm{SG}$ closely follows $\chi N_{g}^\mathrm{eff}$. Numerical simulations of $\Delta_\mathrm{SG}$ calculated using a similar analysis (solid curve) agree qualitatively with experimental values but deviate for larger drive angles. We attribute this to the additional dephasing mechanism described earlier, which would weaken the interaction strength and therefore reduce the size of the gap.

\begin{figure}
    \includegraphics[keepaspectratio, width=\columnwidth]{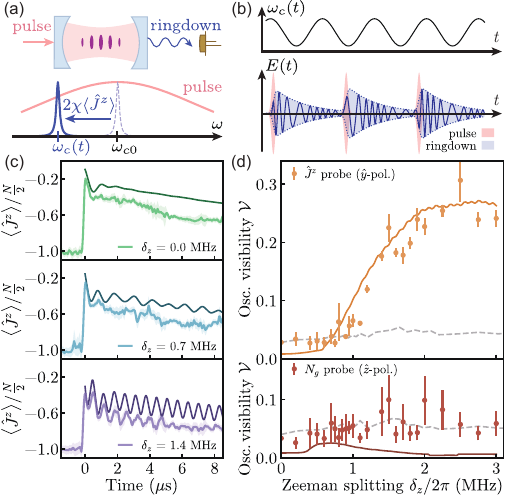}
    \caption{Directly measuring excited state population transfer.
    (a) We probe the cavity resonance by sending light pulses at the cavity and detecting the transmitted light (``ringdown''). The pulses resonantly excite the cavity over a large range of $\omega_{c}(t) = \omega_{c0} - 2\chi\langle \hat{J}^{z} \rangle$.
    (b) The intracavity field adiabatically follows $\omega_c(t)$, which can be measured in ringdown to infer $\langle\hat{J}^{z}\rangle$.
    (c) Time traces of $\langle \hat{J}^{z} \rangle$ (colored curves) with $\pm 1\sigma$ bounds for different $\delta_z$, compared with numerical simulations (thinner dark curves). $\langle \hat{J}^{z} \rangle$ is estimated in $50$~ns time bins and then smoothed with a Gaussian filter ($\tau_\mathrm{smooth} = 50~$ns).
    (d) Top: $\langle\hat{J}^{z}\rangle$ oscillation visibility (circles) from $t = 0.5~\mu$s to $t = 5~\mu$s, relative to a baseline at $\langle\hat{J}^{z}\rangle=-N/2$, alongside numerical simulations (solid curve) and a noise floor (dashed line) inferred from data at $\delta_z/2\pi = 0$~MHz. The noise floor is not flat because calculating visibility involves normalizing each data point separately. Bottom: a similar analysis for an orthogonal $\hat{z}$-polarized probe, which is sensitive to $N_g$ instead of $\langle\hat{J}^{z}\rangle$. The noise floor is consistent with the top plot to aid in a visual comparison of the two probes.
    } \label{fig4}
\end{figure}

We also directly probe internal state population dynamics using a nondestructive measurement of the atomic inversion $\langle\hat{J}^{z}\rangle$. In addition to the atomic Hamiltonian in Eq.~\ref{eq:ham_a}, the total effective Hamiltonian includes an atom-cavity contribution \cite{schleiersmith_2010_pra_vuletic}:
\begin{equation}
    \hat{H}_{\mathrm{ac}}/\hbar = \big(\Delta_c - 2\chi \hat{J}^z\big) \hat{a}^\dagger \hat{a},
    \label{eq:ham_ac}
\end{equation}
where $\hat{a}$ is the annihilation operator for the $\hat{y}$-polarized cavity mode. The cavity resonance frequency $\omega_c(t)$ then varies in response to $\langle\hat{J}^{z}\rangle$, allowing us to infer the dynamics of $\langle\hat{J}^{z}\rangle$ by measuring $\omega_c(t)$.
Our probe scheme, explained fully in the Supplemental Material \cite{supp}, involves applying $100~$ns laser pulses with center frequency $\omega_{c0}$ every $5~\mu$s, similar to work performed in a resonant atom-cavity system \cite{zhu_1990_prl}.
As illustrated in Fig.~\hyperref[fig4]{\ref{fig4}(a)}, the short pulse length induces Fourier broadening, allowing excitation of the cavity mode even if $\omega_c(t)$ deviates by several MHz. This feature gives our probe a large dynamic range for $\omega_c(t)$. 
After each pulse, the cavity freely rings down, and the leakage ``ringdown'' light exhibits a frequency that adiabatically follows $\omega_c(t)$ (Fig.~\hyperref[fig4]{\ref{fig4}(b)}), which we measure to resolve fast changes in $\omega_c(t)$.

Fig.~\hyperref[fig4]{\ref{fig4}(c)} shows time dynamics of $\langle\hat{J}^{z}\rangle$ extracted from our probe. Before the sequence begins, $\langle \hat{J}^{z} \rangle = -N/2$. An initialization pulse at $t=0$ with drive angle $\theta_0 = 0.5\pi$ rapidly increases $\langle \hat{J}^{z} \rangle$.
Afterwards, we observe oscillations in $\langle \hat{J}^{z} \rangle$ in some datasets. For smaller $\delta_z$, these oscillations are hard to resolve above the probe noise floor, whereas larger $\delta_z$ sets display more prominent oscillations.
Assuming $N_{g}$ does not change significantly, oscillations in $\langle \hat{J}^{z} \rangle/\tfrac{N}{2} = (N_{b} - N_{g})/N$ reflect transfer between $\ket{b}$ and $\ket{d}$.
We characterize this process by plotting the oscillation visibility for different $\delta_z$ relative to a baseline at $\langle\hat{J}^{z}\rangle=-N/2$ (orange points in Fig.~\hyperref[fig4]{\ref{fig4}(d)}).
Below $\delta_z/2\pi = 1$~MHz, the lack of visibile oscillations above the noise floor indicates poor population transfer. Above this point, however, the visibility sharply rises and plateaus above $2$~MHz, suggestive of large oscillations between $\ket{b}$ and $\ket{d}$.
The experimental data agrees well with numerical simulations (solid lines).

To confirm that the oscillations in $\langle J^{z} \rangle$ arise from dynamics in $N_b$ and not $N_g$, we also use a $\hat{z}$\nobreakdash-polarized cavity probe (red points in Fig.~\hyperref[fig4]{\ref{fig4}(d)}). In analogy with the $\hat{y}$-polarized probe, this probe measures oscillations of $(N_0 - N_g)/N$, where $N_0$ is the population in $\ket{^{3}P_{1}, m_j=0}$. Since we nominally do not excite this state, we assume $N_0$ is small, such that this probe solely estimates changes in $N_g$. We do not measure an oscillation visibility significantly above the noise floor, so we conclude that the observed oscillations in $\langle\hat{J}^{z}\rangle$ primarily represent population transfer between $\ket{b}$ and $\ket{d}$.

In this work, we have clarified the difference between two distinct many-body gaps, $\Delta_\mathrm{BCS}$ and $\Delta_\mathrm{SG}$, using a clean emulation of the BCS model where the results can be clearly interpreted.
In the future, we could further explore the rich features of BCS superfluidity, such as higher interaction orders, using additional features in our system \cite{shankar_2022_prxquantum}.
This experiment also shows the potential promise in leveraging multiple internal states for exploring more complex physics. Future directions could use the high degeneracy of nuclear spin states in $^{87}$Sr to engineer correlated hopping processes between atoms along the spin manifold \cite{chu_2023_prr_amr} or seed parallel superradiance processes that destructively interfere to form entangled dark states, which could generate optical spin squeezing of interest for optical clocks and matter-wave interferometers \cite{orioli_2022_prx_amr,sundar_2023_prl_rey,sundar_2024_prl_amr}.

\begin{acknowledgments}
We thank Junyu Lin, Maya Miklos, and Jun Ye for stimulating discussions and a careful reading of the manuscript. This material is based upon work supported by the U.S. Department of Energy, Office of Science, National Quantum Information Science Research Centers, Quantum Systems Accelerator. We acknowledge additional funding support from the VBFF, the National Science Foundation under Grant Numbers PFC PHY-2317149  (Physics Frontier Center), OMA-2016244 (QLCI Q-SEnSE) and NIST. D.B. was supported by the Simons collaboration on Ultra-Quantum Matter (UQM)  which is funded by grants from the Simons Foundation (Grant No. 651440), and acknowledges the hospitality of the KITP while parts of this work were completed.
\end{acknowledgments}

\putbib
\end{bibunit}

\clearpage

\title{Supplementary Information: Time-resolved pairing gap spectroscopy in a quantum simulator of fermionic superfluidity inside an optical cavity}

\maketitle
\onecolumngrid

\renewcommand{\thefigure}{S\arabic{figure}}
\setcounter{figure}{0}
\renewcommand{\theequation}{S\arabic{equation}}
\setcounter{equation}{0}
\setcounter{section}{0}
\setcounter{page}{1}
\setcounter{secnumdepth}{2}

\begin{bibunit}
\section{Theory model} \label{sec:model}
We consider cavity-mediated interactions in an effective three-level system using ${}^{88}$Sr atoms in optical cavity. The cavity axis is in the $\hat{x}$ direction, the quantization axis of atoms is in the $\hat{z}$ direction, and the electric field polarization of the cavity mode is in the $\hat{y}$ direction (see Fig.~1 in the main text). 
Here we use $|g\rangle$ to denote the $|{}^1S_0,m=0\rangle$ state of ${}^{88}$Sr atoms, and $|\pm 1\rangle$ for the $|{}^3P_1,m=\pm 1\rangle$ states.
As we discussed below, we can restrict the dynamics in the subspace spanned by $\{|g\rangle,|-1\rangle,|+1\rangle\}$, and consider each atom as an effective three-level system. 

The $\hat{y}$-polarized cavity mode with frequency $\omega_c$ is coupled to the atomic transition (frequency $\omega_a$) $|g\rangle\rightarrow (|+1\rangle+|-1\rangle)/\sqrt{2}$,
leading to the following atom-light coupling Hamiltonian,
\begin{equation}
    \hat{H}_{\rm AL}/\hbar=g_c\sum_j\bigg[i\eta_j \hat{a} \frac{|+1\rangle_j\langle g|+|-1\rangle_j\langle g|}{\sqrt{2}}+h.c.\bigg],
\end{equation}
where $j$ is the atomic index, $\hat{a}$ is the annihilation operator of the cavity mode, $g_c$ is the peak atom-cavity coupling strength, and $\eta_j$ are dimensionless numbers of order $1$ that describe the spatial inhomogeneity in the couplings.
In a standing wave cavity, we have $\eta_j=\cos(n_j\varphi)$ with $\varphi=\pi\lambda_L/\lambda_c$, where $\lambda_c$ is the cavity wavelength, $\lambda_L$ is the lattice wavelength, and $n_j$ is the lattice site index.

Additionally, we consider the Zeeman shift $\delta_z$ generated by the external magnetic field,
\begin{equation}
    \hat{H}_{\rm Z}/\hbar=\frac{\delta_z}{2}\sum_j \bigg[|+1\rangle_j\langle +1|-|-1\rangle_j\langle -1|\bigg],
\end{equation}
as well as the tensor light shift generated by the optical lattice,
\begin{equation}
    \hat{H}_{\rm T}/\hbar= -\sum_j \delta_{\mathrm{T},j} |g\rangle_j\langle g|.
\end{equation}
The inhomogeneity in the tensor light shift is originated by the different lattice intensity for different radial modes.

We consider the pump laser has frequency $\omega_p$. In the rotating frame of the pump laser, the full Hamiltonian of this system can be written as
\begin{equation} \label{eq:ham_full}
    \begin{aligned}
    \hat{H}/\hbar &= g_c\sum_j\bigg[i\eta_j \hat{a} \frac{|+1\rangle_j\langle g|+|-1\rangle_j\langle g|}{\sqrt{2}}+h.c.\bigg] + \frac{\delta_z}{2}\sum_j \bigg[|+1\rangle_j\langle +1|-|-1\rangle_j\langle -1|\bigg] - \sum_j \epsilon_j |g\rangle_j\langle g|\\
    &+\Delta_c \hat{a}^{\dag}\hat{a} + \zeta \hat{a}^{\dag} + \zeta^{*} \hat{a},
    \end{aligned}
\end{equation}
where $\Delta_c=\omega_c-\omega_p$, $\epsilon_j=\omega_a-\omega_p+\delta_{T,j}$, and $\zeta$ is the amplitude of the injected field in the cavity.

In this system, it is convenient to choose the atomic basis as bright state $|b\rangle$ (can couple to cavity photon), dark state $|d\rangle$ (cannot couple to cavity photon) and ground state $|g\rangle$ for each atom,
\begin{equation}
    |b\rangle=\frac{|1,1\rangle+|1,-1\rangle}{\sqrt{2}}, \quad |d\rangle=\frac{|1,1\rangle-|1,-1\rangle}{\sqrt{2}}.
\end{equation}
One can define operators $\hat{S}_{\mu\nu,j}=|\mu\rangle_j\langle\nu|$ with $\mu,\nu=\{b,d,g\}$. In terms of these operators, the full Hamiltonian becomes
\begin{equation}
    \hat{H}/\hbar=\frac{\delta_z}{2}\sum_j(\hat{S}_{bd,j}+\hat{S}_{db,j})-\sum_j\epsilon_j\hat{S}_{gg,j}+g_c\sum_j(i\eta_j \hat{S}_{bg,j}\hat{a}-i\eta^{*}_j\hat{S}_{gb,j}\hat{a}^{\dag})+\Delta_c\hat{a}^{\dag}\hat{a}+\zeta \hat{a}^{\dag}+\zeta^{*}\hat{a}.
\end{equation}

In addition to the Hamiltonian dynamics, we also consider the dissipation processes such as cavity loss with a rate $\kappa$, and spontaneous emission with a rate $\gamma$, so the dynamics of this system can be described by the following Lindblad master equation,
\begin{equation}
    \frac{\mathrm{d}}{\mathrm{d}t}\hat{\rho}=-\frac{i}{\hbar}[\hat{H},\hat{\rho}]+\bigg[\hat{L}_{\mathrm{cav}}\hat{\rho} \hat{L}_{\mathrm{cav}}^{\dag}-\frac{1}{2}\{\hat{L}_{\mathrm{cav}}^{\dag}\hat{L}_{\mathrm{cav}},\hat{\rho}\}\bigg]+\sum_{j}\bigg[\hat{L}_{b,j}\hat{\rho} \hat{L}_{b,j}^{\dag}-\frac{1}{2}\{\hat{L}_{b,j}^{\dag}\hat{L}_{b,j},\hat{\rho}\}\bigg]+\sum_{j}\bigg[\hat{L}_{d,j}\hat{\rho} \hat{L}_{d,j}^{\dag}-\frac{1}{2}\{\hat{L}_{d,j}^{\dag}\hat{L}_{d,j},\hat{\rho}\}\bigg],
\end{equation}
where jump operator for cavity loss is given by
\begin{equation}
    \hat{L}_{\mathrm{cav}}=\sqrt{\kappa}\hat{a},
\end{equation}
and the single-particle jump operators for spontaneous emission are given by
\begin{equation}
    \hat{L}_{b,j}=\sqrt{\gamma}\hat{S}_{gb,j}, \quad \hat{L}_{d,j}=\sqrt{\gamma}\hat{S}_{gd,j}.
\end{equation}

\subsection{Adiabatic elimination of cavity photon} \label{sec:model/adiabatic}

In experiment, $\Delta_c$ is the largest frequency scale ($\Delta_c \gg g_c\sqrt{N}$), so we can adiabatically eliminate the cavity photons and obtain an atom-only Hamiltonian. First we expand the annihilation operator of the cavity mode $\hat{a}$ as a sum of coherent state amplitude $\alpha$ (c-number) and quantum fluctuation $\hat{b}$,
\begin{equation}
    \hat{a}=\alpha+\hat{b},\quad \alpha=\frac{-\zeta}{\Delta_c-i\kappa/2},
\end{equation}
so the Hamiltonian becomes
\begin{equation}
    \begin{aligned}
    \hat{H}/\hbar&=\frac{\delta_z}{2}\sum_j(\hat{S}_{bd,j}+\hat{S}_{db,j})-\sum_j\epsilon_j\hat{S}_{gg,j}+\frac{1}{2}\sum_ji(\Omega\eta_j \hat{S}_{bg,j}-\Omega^{*}\eta^{*}_j\hat{S}_{gb,j})\\
    &+g_c\sum_j(i\eta_j \hat{S}_{bg,j}\hat{b}-i\eta^{*}_j\hat{S}_{gb,j}\hat{b}^{\dag})+\Delta_c\hat{b}^{\dag}\hat{b},
    \end{aligned}
\end{equation}
where $\Omega=-2g_c\zeta/(\Delta_c-i\kappa/2)$. The jump operator for cavity loss becomes
\begin{equation}
    \hat{L}_{\mathrm{cav}}=\sqrt{\kappa}\hat{b}.
\end{equation}
Then we follow the Reiter-S{\o}rensen approach \cite{reiter2012} to obtain the effective Hamiltonian and jump operator,
\begin{equation} \label{eq:ham_eff}
    \hat{H}_{\mathrm{eff}}=\frac{\delta_z}{2}\sum_j(\hat{S}_{bd,j}+\hat{S}_{db,j})-\sum_j\epsilon_j\hat{S}_{gg,j}+\frac{1}{2}\sum_ji(\Omega\eta_j \hat{S}_{bg,j}-\Omega^{*}\eta^{*}_j\hat{S}_{gb,j})-\chi\sum_{jk}\eta_j\eta^{*}_k\hat{S}_{bg,j}\hat{S}_{gb,k},
\end{equation}
\begin{equation}
    \hat{L}_{\mathrm{col}}=\sqrt{\Gamma}\sum_j\eta_j^{*}\hat{S}_{gb,j},
\end{equation}
where 
\begin{equation}
    \chi=\frac{g_c^2\Delta_c}{\Delta_c^2+\kappa^2/4}, \quad \Gamma=\frac{g_c^2\kappa}{\Delta_c^2+\kappa^2/4}.
\end{equation}
The effective Hamiltonian contains the exchange interaction between $|g\rangle$ and $|b\rangle$ states, and the effective jump operator generates superradiant decay of the $|b\rangle$ state.
The effective master equation of this system is thus given by
\begin{equation}
    \frac{\mathrm{d}}{\mathrm{d}t}\hat{\rho}=-\frac{i}{\hbar}[\hat{H}_{\mathrm{eff}},\hat{\rho}]+\bigg[\hat{L}_{\mathrm{col}}\hat{\rho} \hat{L}_{\mathrm{col}}^{\dag}-\frac{1}{2}\{\hat{L}_{\mathrm{col}}^{\dag}\hat{L}_{\mathrm{col}},\hat{\rho}\}\bigg]+\sum_{j}\bigg[\hat{L}_{b,j}\hat{\rho} \hat{L}_{b,j}^{\dag}-\frac{1}{2}\{\hat{L}_{b,j}^{\dag}\hat{L}_{b,j},\hat{\rho}\}\bigg]+\sum_{j}\bigg[\hat{L}_{d,j}\hat{\rho} \hat{L}_{d,j}^{\dag}-\frac{1}{2}\{\hat{L}_{d,j}^{\dag}\hat{L}_{d,j},\hat{\rho}\}\bigg].
\end{equation}

In this case, it's convenient to calculate the Heisenberg equation of motion for operators $\hat{S}_{\mu\nu,j}$ in large $N$ limit, and take the expectation value to get mean-field equations (we remove the hat of the operators to represent expectation values). Using the commutation relations $[\hat{S}_{\mu\nu,j},\hat{S}_{\rho\sigma,k}]=(\hat{S}_{\mu\sigma,j}\delta_{\nu\rho}-\hat{S}_{\rho\nu,j}\delta_{\sigma\mu})\delta_{jk}$, we have
\begin{equation}
    \begin{gathered}
        \begin{aligned}
        \frac{\mathrm{d}}{\mathrm{d}t}S_{gg,j}&=-\frac{1}{2}(\Omega\eta_jS_{bg,j}+\Omega^{*}\eta_j^{*}S_{gb,j})-i\chi\sum_k(\eta_j\eta_k^{*}S_{gb,k}S_{bg,j}-\eta_j^{*}\eta_kS_{bg,k}S_{gb,j})\\
        &+\frac{\Gamma}{2}\sum_k(\eta_j\eta_k^{*}S_{gb,k}S_{bg,j}+\eta_j^{*}\eta_kS_{bg,k}S_{gb,j})+\gamma(S_{bb,j}+S_{dd,j}),
        \end{aligned}\\
        \begin{aligned}
        \frac{\mathrm{d}}{\mathrm{d}t}S_{bb,j}&=i\frac{\delta_z}{2}(-S_{bd,j}+S_{db,j})+\frac{1}{2}(\Omega\eta_jS_{bg,j}+\Omega^{*}\eta_j^{*}S_{gb,j})+i\chi\sum_k(\eta_j\eta_k^{*}S_{gb,k}S_{bg,j}-\eta_j^{*}\eta_kS_{bg,k}S_{gb,j})\\
        &-\frac{\Gamma}{2}\sum_k(\eta_j\eta_k^{*}S_{gb,k}S_{bg,j}+\eta_j^{*}\eta_kS_{bg,k}S_{gb,j})-\gamma S_{bb,j},
        \end{aligned}\\
        \begin{aligned}
        \frac{\mathrm{d}}{\mathrm{d}t}S_{gb,j}&=-i\frac{\delta_z}{2}S_{gd,j}-\frac{\Omega}{2}\eta_j(S_{bb,j}-S_{gg,j})-i\chi\sum_k\eta_j\eta_k^{*}S_{gb,k}(S_{bb,j}-S_{gg,j})-i\epsilon_jS_{gb,j}\\
        &+\frac{\Gamma}{2}\sum_k\eta_j\eta_k^{*}S_{gb,k}(S_{bb,j}-S_{gg,j})-\frac{\gamma}{2}S_{gb,j},
        \end{aligned}\\
        \frac{\mathrm{d}}{\mathrm{d}t}S_{gd,j}=-i\frac{\delta_z}{2}S_{gb,j}-\frac{\Omega}{2}\eta_jS_{bd,j}-i\chi\sum_k\eta_j\eta_k^{*}S_{gb,k}S_{bd,j}-i\epsilon_jS_{gd,j}+\frac{\Gamma}{2}\sum_k\eta_j\eta_k^{*}S_{gb,k}S_{bd,j}-\frac{\gamma}{2}S_{gd,j},\\
        \frac{\mathrm{d}}{\mathrm{d}t}S_{bd,j}=i\frac{\delta_z}{2}(S_{dd,j}-S_{bb,j})+\frac{\Omega^{*}}{2}\eta_j^{*}S_{gd,j}-i\chi\sum_k\eta_j^{*}\eta_kS_{bg,k}S_{gd,j}-\frac{\Gamma}{2}\sum_k\eta_j^{*}\eta_kS_{bg,k}S_{gd,j}-\gamma S_{bd,j}.
    \end{gathered}
\end{equation}
Note that the total atom number is conserved in our model, we do not need to include the differential equation for $S_{dd,j}$ because $S_{dd,j}=1-S_{gg,j}-S_{bb,j}$. For the numerical simulation, we first turn on $\Omega$ to prepare the corresponding initial states, and then turn off $\Omega$ and let the system evolve.

\subsection{Mapping to radio-frequency spectroscopy in ultracold fermionic systems}
We consider the following fermionic system with three internal states: $|1\rangle$, $|2\rangle$, $|3\rangle$. We assume the s-wave interaction between $|1\rangle$ and $|2\rangle$ is much larger than the others (close to Feshbach resonance), so we can neglect the interactions related to $|3\rangle$ and only consider BCS pairing between $|1\rangle$ and $|2\rangle$ states. We consider a radio-frequency (RF) drive (no momentum transfer) between $|1\rangle$ and $|3\rangle$ states.
RF spectroscopy experiment is to scan the frequency of the RF drive, while in our case we fixed the frequency resonating with the bare transition frequency between $|1\rangle$ and $|3\rangle$ states. The Hamiltonian of this system can be written as
\begin{equation}
    \hat{H}/\hbar=\sum_{\mathbf{k}}\sum_{\sigma=\{1,2,3\}}\epsilon_{\mathbf{k}}\hat{c}_{\mathbf{k},\sigma}^{\dag}\hat{c}_{\mathbf{k},\sigma}-\chi\sum_{\mathbf{k}\mathbf{k}'}\hat{c}_{\mathbf{k},1}^{\dag}\hat{c}_{-\mathbf{k},2}^{\dag}\hat{c}_{-\mathbf{k}',2}\hat{c}_{\mathbf{k}',1}+\frac{\delta_z}{2}\sum_{\mathbf{k}}(\hat{c}_{\mathbf{k},3}^{\dag}\hat{c}_{\mathbf{k},1}+h.c.).
\end{equation}
Similar to the Anderson pseudospin mapping \cite{Anderson1958}, One can define a complete basis for this Hamiltonian as follows,
\begin{equation}
    \begin{gathered}
    |g\rangle_{\mathbf{k}}=|n_{\mathbf{k},1}=0,n_{-\mathbf{k},2}=0,n_{\mathbf{k},3}=0\rangle,\\
    |b\rangle_{\mathbf{k}}=|n_{\mathbf{k},1}=1,n_{-\mathbf{k},2}=1,n_{\mathbf{k},3}=0\rangle,\\
    |d\rangle_{\mathbf{k}}=|n_{\mathbf{k},1}=0,n_{-\mathbf{k},2}=1,n_{\mathbf{k},3}=1\rangle.\\
    \end{gathered}
\end{equation}
So we can rewrite this Hamiltonian using operators $\hat{S}_{\mu\nu,\mathbf{k}}=|\mu\rangle_\mathbf{k}\langle\nu|$ with $\mu,\nu=\{b,d,g\}$, which gives
\begin{equation}
    \hat{H}/\hbar=-\sum_{\mathbf{k}}\epsilon_{\mathbf{k}}\hat{S}_{gg,\mathbf{k}}-\chi\sum_{\mathbf{k}\mathbf{k}'}\hat{S}_{bg,\mathbf{k}}\hat{S}_{gb,\mathbf{k}'}+\frac{\delta_z}{2}\sum_{\mathbf{k}}(\hat{S}_{db,\mathbf{k}}+h.c.),
\end{equation}
which agrees with the effective Hamiltonian of our three-level system in cavity.

\section{Analytic discussions} \label{sec:analytics}
For simplicity, we consider analytic discussions in the case of homogeneous atom-light couplings, $\eta_j=1$. We further assume the interaction strength $\chi N$ is much larger than the frequency disorder $\epsilon_j$, so we can drop the $\epsilon_j$ terms and restrict the dynamics in the fully symmetric manifold. We then rewrite the Hamiltonian in terms of Schwinger bosons for the three internal states, i.e. $(\hat{a}_g,\hat{a}_b,\hat{a}_d)$,
\begin{equation}
    \hat{H}=\frac{\delta_z}{2}(\hat{a}^{\dag}_b\hat{a}_d+\hat{a}^{\dag}_d\hat{a}_b)-\chi \hat{a}^{\dag}_b\hat{a}_g\hat{a}^{\dag}_g\hat{a}_b.
    \label{eq:sch}
\end{equation}
Here we set $\Omega=0$ during the time evolution. 
In terms of Schwinger bosons, one can easily calculate the energy shift in RF spectroscopy, which is equivalent to the energy difference between $|b\rangle$ and $|d\rangle$ states.
Define $\hat{N}_{\mu}=\hat{a}_{\mu}^{\dag}\hat{a}_{\mu}$ with $\mu=\{b,d,g\}$, we have $\chi \hat{a}^{\dag}_b\hat{a}_g\hat{a}^{\dag}_g\hat{a}_b=\chi\hat{N}_b(\hat{N}_g+1)$.
In the large-$N$ limit, we have $E_b-E_d\approx \chi N_g$.

\subsection{Mean-field solutions in the ideal case}
The mean-field approximation for Eq.~(\ref{eq:sch}) is to derive the Heisenberg equation of motion for $(\hat{a}_g,\hat{a}_b,\hat{a}_d)$, and then replace them by $c$-numbers: $(\hat{a}_g,\hat{a}_b,\hat{a}_d)\sim \sqrt{N}(v_g,v_b,v_d)^T$, where $|v_g|^2+|v_b|^2+|v_d|^2=1$. So the mean-field equations in terms of $v_g,v_b,v_d$ takes the following form,
\begin{equation}
    \begin{gathered}
        \dot{v}_g=iN\chi(v_b^{*}v_b)v_0,\\
        \dot{v}_b=-i\bigg(\frac{\delta_z}{2}v_d-N\chi(v_g^{*}v_g)v_b\bigg),\\
        \dot{v}_d=-i\frac{\delta_z}{2}v_b.\\
    \end{gathered}
\end{equation}
The first equation implies that $v_g^{*}v_g=N_g/N=\mathrm{const}$. Therefore, we can simplify the equations for $v_b$ and $v_d$ as follows,
\begin{equation}
    \begin{gathered}
        \dot{v}_b=-i\bigg(\frac{\delta_z}{2}v_d-N_g\chi v_b\bigg),\\
        \dot{v}_d=-i\frac{\delta_z}{2}v_b,\\
    \end{gathered}
\end{equation}
which are linear and can be solved analytically. So the general solution is given by
\begin{equation}
    \begin{gathered}
        v_g(t)=v_g(0)\exp\bigg[iN\chi\int_0^t v_b^{*}(t')v_b(t')\mathrm{d}t'\bigg],\\
        v_b(t)=\bigg[v_b(0)\cos\bigg(\frac{\omega t}{2}\bigg)+iv_b(0)\frac{N_g\chi}{\omega}\sin\bigg(\frac{\omega t}{2}\bigg)-iv_d(0)\frac{\delta_z}{\omega}\sin\bigg(\frac{\omega t}{2}\bigg)\bigg]\exp\bigg(i\frac{N_g\chi}{2}t\bigg),\\
        v_d(t)=\bigg[v_d(0)\cos\bigg(\frac{\omega t}{2}\bigg)-iv_d(0)\frac{N_g\chi}{\omega}\sin\bigg(\frac{\omega t}{2}\bigg)-iv_b(0)\frac{\delta_z}{\omega}\sin\bigg(\frac{\omega t}{2}\bigg)\bigg]\exp\bigg(i\frac{N_g\chi}{2}t\bigg),\\
    \end{gathered}
\end{equation}
where
\begin{equation}
    \omega=\sqrt{(N_g\chi)^2+\delta_z^2}.
\end{equation}

So we can calculate the value of relevant experimental observables based on the general solution above. The amplitude of BCS order parameter is given by $|\Delta_{\rm BCS}(t)|/\chi N=|v_b^{*}(t)v_g(t)|$, and the dark state population is given by $N_d(t)=v_d^{*}(t)v_d(t)$. In both cases, they are oscillating with frequency $\omega$. 

\newpage

\section{Experimental details} \label{sec:experiment}
\subsection{The QND Hamiltonian} \label{sec:experiment/qnd}
Eq.~(\ref{eq:ham_full}) provides a full description of relevant features of the system. However, the effective Hamiltonian derived in Eq.~(\ref{eq:ham_eff}) does not include the additional atom-cavity Hamiltonian term introduced in the main text, which describes the key physics behind our nondestructive cavity probe. Where does the additional term come from? 

The effective Hamiltonian description in Sec.~\ref{sec:model/adiabatic} relies on an adiabatic elimination step that assumes all relevant dynamics occurs close to DC in the rotating frame of the pump laser, which is assumed to be resonant with the atomic transition. In this picture, because the cavity resonance frequency is sufficiently far-detuned, we conclude that the cavity field responds to the atoms ``driving'' the mode very quickly and thus can adiabatically eliminate the field. However, in the presence of an applied cavity probe laser close to resonance with the cavity (and not the atoms), the system can respond in two frequency bands, complicating this picture.

We describe the pump field $\zeta$ using two terms: $\zeta = \zeta^{(a)} + \zeta^{(c)} e^{-i \Delta_c t}$, near resonance with the atomic transition and cavity mode respectively. The two fields can vary in time, but we can cleanly separate the responses if we assume that $\lvert\dot{\zeta}^{(\sigma)}\rvert \ll \Delta_{c} \lvert\zeta^{(\sigma)}\rvert$, such that the Fourier transform of the field exhibits two well-separated regions centered at DC and $\Delta_c$. In experimental terms, $\zeta^{(a)}$ represents the initial laser drive that excites the atoms. $\zeta^{(c)}$ describes the cavity probe, which is described in detail in the next section. We can characterize the mean-field response of the system to these two separate drives by setting up the following ansatz:
\begin{align}
\begin{split}
    \langle \hat{a} \rangle = a &\coloneqq a^{(a)} + a^{(c)} e^{-i \Delta_c t} \\
    \langle \hat{S}_{gb,j} \rangle = S_{gb,j} &\coloneqq S_{gb,j}^{(a)} + S_{gb,j}^{(c)} e^{-i \Delta_c t} \\
    \langle \hat{S}_{gb,j} \rangle = S_{gd,j} &\coloneqq S_{gd,j}^{(a)} + S_{gd,j}^{(c)} e^{-i \Delta_c t}
\end{split}
\end{align}
The other atomic observables ($\hat{S}_{bd,j}$, $\hat{S}_{bb,j}$, $\hat{S}_{gg,j}$, and $\hat{S}_{dd,j}$) do not represent optical coherences and thus do not exhibit a response in the two (optical frequency) bands, which we will call the ``atomic band'' and the ``cavity band.'' In principle, the responses to the two bands can be coupled because the atomic degrees of freedom are highly nonlinear. However, if we assume any dynamics from the cavity band are sufficiently weak and that $\Delta_c$ represents the largest dynamical frequency scale in the system, to leading order the responses are decoupled. Experimentally, this represents the limit taken by quantum nondemolition (QND) probes, which are designed to extract information without significantly perturbing the atoms.

In the QND limit, the atomic band responses are well-described by Sec.~\ref{sec:model/adiabatic}. We can calculate the mean-field cavity band responses by applying the Heisenberg equation of motion to Eq.~(\ref{eq:ham_full}) and separating out the frequency response at $\Delta_c$. This yields the following set of equations:
\begin{align}
\begin{split}
    \frac{\mathrm{d}}{\mathrm{d}t} a^{(c)} &= -\frac{\kappa}{2} a^{(c)} - g_c \sum_{j} \eta_{j} S_{gb,j}^{(c)} - i \zeta^{(c)} \\
    \frac{\mathrm{d}}{\mathrm{d}t} S_{gb,j}^{(c)} &= \big[i(\Delta_c - \epsilon_{j}) - \frac{\gamma}{2}\big] S_{gb,j}^{(c)} - i \frac{\delta_z}{2} S_{gd,j}^{(c)} - g_c \eta_{j} a^{(c)} (S_{bb,j} - S_{gg,j}) \\
    \frac{\mathrm{d}}{\mathrm{d}t} S_{gd,j}^{(c)} &= \big[i(\Delta_c - \epsilon_{j}) - \frac{\gamma}{2}\big] S_{gd,j}^{(c)} - i \frac{\delta_z}{2} S_{gb,j}^{(c)} - g_c \eta_{j} a^{(c)} S_{bd,j},
\end{split}
\end{align}
Note that, unlike with the atomic band, equations in the cavity band can have a macroscopic cavity mode population, and instead the atomic coherences adiabatically follow the cavity since they are off resonance by roughly $\Delta_c$. Therefore, we can set the time derivatives of the atomic coherences to 0. Solving the coupled equations to leading order in $\Delta_c^{-1}$ yields:
\begin{align}
\begin{split}
    S_{gb,j}^{(c)} &= \frac{g_c \eta_{j} a^{(c)}}{
    \big[i(\Delta_c - \epsilon_{j}) - \frac{\gamma}{2}\big]^2 + \tfrac{\delta_z^2}{4}}
    \left( i\frac{\delta_z}{2} S_{bd,j} + \big[i(\Delta_c - \epsilon_{j}) - \frac{\gamma}{2}\big] (S_{bb,j} - S_{gg,j})\right) \\
    &= -i \frac{g_c}{\Delta_c} \eta_j a^{(c)} (S_{bb,j} - S_{gg,j}) + \mathcal{O}(\Delta_{c}^{-2}) \\
    S_{gd,j}^{(c)} &= \frac{1}{\big[i(\Delta_c - \epsilon_{j}) - \frac{\gamma}{2}\big]}
    \left( i\frac{\delta_z}{2} S_{gb,j}^{(c)} + g_c \eta_j a^{(c)} S_{bd,j} \right) \\
    &= -i \frac{g_c}{\Delta_c} \eta_j a^{(c)} S_{bd,j} + \mathcal{O}(\Delta_{c}^{-2}).
\end{split}
\end{align}

We can then eliminate the atomic observables from the equation of motion for the classical field $a^{(c)}$ to obtain:
\begin{align} \label{eq:cavity_field}
\begin{split}
    \frac{\mathrm{d}}{\mathrm{d}t} a^{(c)} &= -\frac{\kappa}{2} a^{(c)} + i \frac{g_{c}^2}{\Delta_c} \sum_{j} \eta_{j}^2 (S_{bb,j} - S_{gg,j}) a^{(c)} - i \zeta^{(c)} \\
    &= \big[2i\chi_{0} \sum_{j} \eta_{j}^2 J_{j}^{z} - \frac{\kappa}{2} \big] a^{(c)} - i \zeta^{(c)},
\end{split}
\end{align}
where $\chi_0 \approx g_{c}^2/\Delta_c$ is the spin-exchange interaction strength (for peak couplers) taken to the limit $\Delta_c \gg \kappa/2$, and $J_{j}^{z} = \tfrac{1}{2}(S_{bb,j} - S_{gg,j})$ is the inversion for atom $j$ along the bright-ground transition. We see that, at least at the mean-field level, the presence of atoms in the cavity acts as an effective frequency shift. Experimental intuition comes from recognizing that the near-resonant atoms act as a gas with an inversion-dependent index of refraction, which changes the cavity resonance frequency.

Using adiabatic elimination techniques (for instance, by applying the Reiter-S{\o}rensen approach \cite{reiter2012} for each $m_J$ subspace separately), it is possible to promote this mean-field intuition into an effective QND Hamiltonian for the cavity field:
\begin{equation} \label{eq:ham_qnd}
    \hat{H}_\mathrm{QND} = \left( \Delta_c - 2\chi_0 \sum_{j} \eta_{j}^2 \hat{J}_{j}^{z} \right)
    \hat{a}^{\dagger} \hat{a}.
\end{equation}
We can follow definitions from the Supplemental Material of a previous work \cite{norcia_2018_science_jkt} and define an effective ``weighted'' inversion operator of the form 
\begin{equation}
    \hat{J}^{z \prime} \coloneqq \frac{\sum_{j} \eta_{j}^2 \hat{J}_{j}^{z}}{\tfrac{1}{N}\sum_{j} \eta_{j}^2},
\end{equation}
defined such that $\hat{J}^{z \prime}$ takes values between $\pm N/2$. Combined with an rms interaction strength $\chi^{\prime} = \chi_0 \big( \tfrac{1}{N} \sum_{j} \eta_{j}^2 \big)$, Eq.~(\ref{eq:ham_qnd}) simplifies to $\hat{H}_\mathrm{QND} = (\Delta_c - 2\chi^{\prime} \hat{J}^{z \prime}) \hat{a}^{\dagger} \hat{a}$, which is the form presented in the main text.
In the experiment, the atoms exhibit couplings characterized by $\eta_{j} = \cos(\varphi_{j})$ with uniformly distributed phases $\varphi_{j} \in [0,2\pi)$. In this case, the weighted definitions evaluate to $\hat{J}^{z \prime} = 2 \sum_{j} \eta_{j}^2 \hat{J}_{j}^{z}$, $\chi^{\prime} = \chi_0 /2$. 
Note that both $\chi$ and $\hat{J}^{z}$ in the main text always refer to the primed quantities defined here, which reflects weighting by cavity coupling.

\subsection{The cavity probe} \label{sec:experiment/probe}
In order to measure the cavity resonance frequency $\omega_{c}(t)$, we apply a series of probe pulses at the cavity by sending RF pulses into a fiber phase modulator to periodically create an FM sideband nominally resonant with the cavity. To shape each RF pulse, we program an arbitrary waveform generator to produce an RF pulse of the form
\begin{equation} \label{eq:cavprobe_pulse}
    V(t) = V_\mathrm{env}(t) \cos(\tilde{\omega} t);\qquad V_\mathrm{env}(t) = V_0 e^{-\tfrac{t^2}{2 \tau^2}};\qquad \tau = 50~\mathrm{ns},
\end{equation}
where $\tilde{\omega} = \omega_{c0} - \omega_{l}$ is the RF frequency necessary to generate a first-order sideband at the unshifted cavity resonance $\omega_{c0}$, given input light with frequency $\omega_{l}$. In our case, $\tilde{\omega}/2\pi = -60~$MHz, which is large enough that $V_\mathrm{env}(t)$ varies slowly compared to $\tilde{\omega}$ ($\lvert\tilde{\omega}\rvert\tau \gg 1$).
The induced FM modulation therefore results in resolved sidebands of light, which follow a Jacobi-Anger expansion:
\begin{align}
\begin{split}
    E(t) &= E_{0} e^{i \omega_{l} t} \exp\Big(i \tfrac{V_\mathrm{env}(t)}{V_\pi} \pi \cos(\tilde{\omega}t)\Big) \\
    &= E_{0} \sum_{n = -\infty}^{\infty} i^n J_{n}\Big(\tfrac{V_\mathrm{env}(t)}{V_\pi} \pi\Big)
    e^{i(\omega_{l} + n\tilde{\omega})t},
\end{split}
\end{align}
where $J_n$ is the $n^\mathrm{th}$-order cylindrical Bessel function, and $V_\pi$ is the half-wave voltage of the fiber phase modulator. In our case, $V_0/V_\pi \approx 0.23$, meaning the modulator is driven close to the linear excitation regime (i.e., $J_{1}(\tfrac{V_\mathrm{env}(t)}{V_\pi}\pi) \approx \tfrac{V_\mathrm{env}(t)}{V_\pi} \tfrac{\pi}{2}$). As a result, we assume that the electric field of the first-order FM sideband roughly inherits the shape described by $V_\mathrm{env}(t)$.

The Fourier spectrum of $V_\mathrm{env}$ is also a Gaussian, with a standard deviation of $\sigma = 1/\tau \approx 2\pi\times 3.18~$MHz and a HWHM of $\sigma\sqrt{2\ln 2} \approx 2\pi\times 3.75~$MHz. This means that if $\omega_{c}(t) - \omega_{0} = 2\pi\times \pm3.75~$MHz at the time of the pulse, the injected electric field will be half as big as it would have been if the cavity were unshifted, resulting in a $50\%$ reduction in signal-to-noise. In other words, the pulse ensures that the probe will have at least $50\%$ of the maximum signal-to-noise over a $7.5~$MHz band centered at $\omega_{c0}$.
Over time, the light inside the cavity will leak out, with a $1/e$ time constant of $2/\kappa \approx 2.1~\mu$s for the electric field. During this time, the cavity field will evolve according to Eq.~(\ref{eq:cavity_field}) with $\zeta^{(c)} = 0$ so long as changes in $\hat{J}^{z \prime}$ occur much more slowly than $\Delta_c$, the scale separating the atomic and cavity bands. Therefore, as the light leaks out of the cavity, its frequency is modulated to match the dynamics of $\hat{J}^{z \prime}$. 

Intuition for why the light field should ``adiabatically follow'' the cavity resonance can be gained in a couple of ways. From a microscopic perspective, changes in $\hat{J}^{z \prime}$ modify the effective index of refraction inside the cavity. Every time a photon passes through the atomic ensemble, it will experience a phase shift in proportion to the index of refraction. Since these passes occur with a rate set by the cavity free spectral range $\omega_\mathrm{FSR}$, so long as $\omega_\mathrm{FSR} \gg \lvert \tfrac{\omega_c'(t)}{\omega_c(t)} \rvert$ we can coarse-grain these shifts and treat them as a frequency shift (i.e., a phase shift per unit time). We can also understand this phenomenon using a Landau-Zener argument. Assuming any couplings between the $\mathrm{TEM}_{00}$ cavity modes and higher-order transverse modes is negligible, the splitting between cavity eigenstates is equal to $\omega_\mathrm{FSR}$, so the probability that the cavity field adiabatically follows the cavity resonance is close to unity as long as $\omega_\mathrm{FSR} \gg \lvert \tfrac{\omega_c'(t)}{\omega_c(t)} \rvert$. In both of these arguments, the presence of an atomic resonance at a frequency $\Delta_c$ away limits the bandwidth of this adiabatic following to $\Delta_c \ll \omega_\mathrm{FSR}$, since changes in $\omega_c(t)$ with non-negligible modulation at $\Delta_c$ will resonantly drive the atoms.

Continuing the description of the probe, we repeat the pulse described above every $5~\mu$s in order to keep the cavity populated with photons. In between pulses, we detect the cavity field in transmission by beating it against a local oscillator and measuring in heterodyne to infer a complex electric field amplitude as a function of time. We then time bin the trace into chunks of length $\Delta t_\mathrm{bin} = 67~$ns (15 bins per $\mu$s, compared with a 60~MS/s acquisition rate for the scope). Within each time bin, we estimate the frequency by calculating the peak Fourier response.

The precision of this probe relies heavily on signal-to-noise considerations. We operate at a power where the maximum intracavity photon number is roughly $M_c \approx 30\times 10^3$. Due to imperfect quantum efficiency, we effectively collect about $40$ photons per time bin per shot of the experiment, which does not allow for a high precision in frequency estimation. To improve this, we repeat the experiment 8 times within a single loading sequence of the experiment, over a time period of roughly 2~ms. In postprocessing, we average the complex electric field phasors from each of these experimental shots and then estimate the frequency, which multiplies the effective photon number by 8. Then, we repeat this loading sequence for a total of 100 times. Since we cannot ensure the phase stability between loading sequences (which lasted $3$~s each), we could not directly average the complex phasors. Instead, we process each loading sequence separately to obtain 100 estimates of the cavity resonance frequency per time bin, which we then average down to increase precision by a factor of $\sqrt{100} = 10$. We empirically find that the spread in frequency estimates is around $200$~kHz at maximum intracavity power, giving us an estimated maximum frequency precision of $20$~kHz. The uncertainty region presented in the main text comes not from this estimated precision but instead from performing a bootstrap resampling on the $100$ frequency estimates per time bin ($n_\mathrm{boot}$ = 100) and calculating the standard deviation over resampled estimates. Finally, since the signal-to-noise decreases exponentially between pulses, we repeat the experiment with the cavity probe pulses applied with a $2.5~\mu$s offset. In postprocessing, we stitch together the two experiments to optimize the signal size in each time bin. We do not observe any distortion of the dynamics by applying the cavity probe pulses at different times.

In principle, we could improve the signal-to-noise of the probe by sending in more photons; however, if the probe becomes too strong it will start to affect the system dynamics (which we do not want). Eq.~(\ref{eq:ham_qnd}) can be interpreted as a shift of the atomic transition frequency by a characteristic amount $\chi^{\prime} \langle \hat{a}^{\dagger} \hat{a} \rangle = \chi^{\prime} M_c$, where $M_c$ is the number of intracavity photons. We want this to be small compared with the frequencies of the Hamiltonian of interest, such as $\chi^{\prime} N \approx 2\pi\times 1~$MHz. For the estimated intracavity photon number $M_c$, we calculate a maximum frequency scale of $\chi^{\prime} M_c \approx (2\pi\times 35~\mathrm{kHz}) \ll \chi^{\prime} N$. The signal-to-noise could be further improved by increasing the quantum efficiency of our system, which is estimated to be around $0.02$. Finally, although we don't believe it to be true in our system, laser frequency noise in the local oscillator could be a limiting factor if it is larger than the shot-noise-limited precision of the probe. Using a laser with narrower linewidth could then also benefit the probe.

\subsection{Effects of inhomogeneous coupling on the spectral gap} \label{sec:experiment/spectralgap}
As outlined in the main text, for a system with homogeneous atom-light coupling, an initial state in the maximally symmetric manifold (i.e., $J=\tfrac{N_b + N_g}{2}$) experiences a spectral gap equal to $\Delta_\mathrm{SG} = \chi N_g$. The presence of inhomogeneous coupling (see Sec.~\ref{sec:model}) disrupts this picture. It is hard to analytically calculate the spectral gap $\Delta_\mathrm{SG}$ in the case of inhomogeneous coupling. Instead, we rely on numerically simulating $\Delta_\mathrm{SG}$ with the same analysis used on the experimental data. Additionally, we compare $\Delta_\mathrm{SG}$ against what we believe is a reasonable analog for $\chi N_g$ for an inhomogeneously coupled system. This analog replaces $N_g$ with an estimate of the ground state population weighted by cavity coupling, which we will call $N_{g}^{\prime}$ (or, in an abuse of notation, just $N_g$ in the main text):
\begin{equation}
    N_{g}^{\prime} \coloneqq \frac{\sum_{j} \eta_{j}^2 N_{g,j}}{\tfrac{1}{N}\sum_{j} \eta_{j}^2},
\end{equation}
normalized such that when all atoms are in the ground state, $N_g^{\prime} = N$.
The intuition for why this seems like a good choice comes from Sec.~\ref{sec:experiment/qnd}, where the weighted atomic inversion $\hat{J}^{z \prime}$ turns out to be the correct observable to determine shifts in the cavity resonance frequency.

\begin{figure}[t]
    \includegraphics{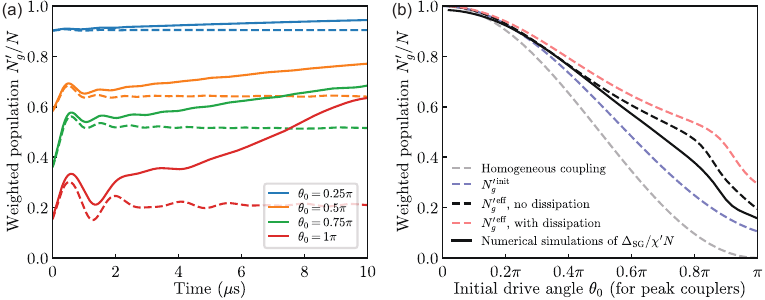}
    \caption{Analysis of the weighted ground state population $N_{g}^{\prime}$.
    (a) Time dynamics after an initial Rabi pulse with drive angle $\theta_0$ for peak couplers. Dashed lines are derived from simulations neglecting single-particle and collective dissipation, and solid lines represent a full numerical simulation of the experiment (including dissipation).
    (b) Comparison of various $N_{g}^{\prime}$ measurements vs. $\theta_0$. The gray dashed trace represents homogeneous coupling, with all other traces assuming uniformly inhomogeneous coupling. The blue dashed trace shows the expected initial-time population according to Eq.~(\ref{eq:Ng_init}). The black and red traces represent $N_{g}^{\prime}$ averaged over the time interval $t=0.5~\mu$s to $t=5~\mu$s (the interval used to measure $\Delta_\mathrm{SG}$), using simulations neglecting and including dissipation, respectively. Finally, the black solid trace is $\Delta_\mathrm{SG}$ computed from numerical simulations with the same method used with the experimental data, normalized to 1 at $\theta_0 = 0$.
    } \label{figS1}
\end{figure}

However, this quantity carries an additional complication: while the unweighted ground state atom number $N_g$ is preserved under the spin-exchange interaction (the number of excitations does not change), $N_g^{\prime}$ is not. This is because when atoms with different cavity coupling strengths exchange excitations, how the excitation is weighted also changes. This can be seen in Fig.~\hyperref[figS1]{\ref{figS1}a}, which shows numerical simulations of $N_{g}^{\prime}$ for several different drive angles $\theta_0$ (defined as the angle experienced by atoms with maximum coupling). We see that $N_{g}^{\prime}$ oscillates for a few $\mu$s after initialization, sometimes drastically changing from the $t=0$ value. Interestingly, in simulations that neglect dissipation (dashed curves), $N_{g}^{\prime}$ appears to quickly approach a steady state, where the system attains a sort of detailed balance in the exchange between classes of atoms with different coupling strengths.
With dissipation (solid curves), $N_{g}^{\prime}$ decays to $N$ at long times.

Fig.~\hyperref[figS1]{\ref{figS1}b} shows how various measurements of $N_{g}^{\prime}$ differ as a function of drive angle $\theta_0$. For homogeneous coupling (gray dashed line), $N_{g}^{\prime} = N \cos(\theta_0/2)^2$ is constant in the absence of dissipation. In the case of inhomogeneous coupling, we assume the atoms uniformly sample a coupling phase $\varphi$, resulting in a coupling strength $\eta(\varphi) = \cos\varphi$. In this case, the weighted ground state population at $t=0$ ($N_{g}^{\prime\,\mathrm{init}}$, blue dashed line) is given by:
\begin{equation} \label{eq:Ng_init}
    N_{g}^{\prime\,\mathrm{init}}(\theta_0) = N \left( 
    \frac{1}{2} + \frac{J_{1}(\theta_0)}{\theta_0} - J_{2}(\theta_0)
    \right).
\end{equation}
In the experiment, we estimate $\Delta_\mathrm{SG}$ by analyzing the oscillation frequency of $\hat{J}^{-}$ within the measurement interval $t \in [0.5~\mu\mathrm{s},5~\mu\mathrm{s}]$. The black dashed line represents the average weighted ground state population in this time interval, which we call the effective ground state population $N_{g}^{\prime\,\mathrm{eff}}$, using numerical simulations without dissipation. By visual inspection of Fig.~\hyperref[figS1]{\ref{figS1}a}, this estimate should be fairly close to the long-time steady-state value fo $N_{g}^{\prime}$. We also plot $N_{g}^{\prime\,\mathrm{eff}}$ using simulations including dissipation (red dashed line). Finally, we compare all of these estimates against $\Delta_\mathrm{SG}$ calculated from numerical simulations and normalized to the same units.

One might be tempted to claim that $\Delta_\mathrm{SG}$ should increase in the presence of dissipation since $N_{g}^{\prime}$ increases over time. While this is true for collective dissipation, it is not necessarily true when single-particle spontaneous emission is the dominant loss process (as is the case in our experiment). This is because the relationship $\Delta_\mathrm{SG} = \chi N_g$ only holds for maximally symmetric states. In general, states with lower angular momentum will also have a smaller $\Delta_\mathrm{SG}$. This can be seen for homogeneously coupled systems by considering a wavefunction of the form $\ket{\Psi_i} = \ket{J, J^{z}}_{gb} \otimes \ket{\psi}_{d}$, which is an eigenstate of the interaction term $\chi \hat{J}^{+} \hat{J}^{-}$ with eigenvalue $E_i = \chi(J(J+1) - J^{z}(J^{z}-1))$. Transferring one atom symmetrically from $\ket{b}$ to $\ket{d}$ transforms the wavefunction to $\ket{\Psi_f} = \ket{J-\tfrac{1}{2}, J^{z}-\tfrac{1}{2}}_{gb} \otimes \ket{\psi^{\prime}}_{d}$, which can be seen using the Schwinger boson formalism introduced in Sec.~\ref{sec:analytics}. This is also an eigenstate of the interaction term with eigenvalue $E_f$, giving a total energy cost equal to
\begin{align}
\begin{split}
    \Delta_\mathrm{SG} = E_i - E_f &= \chi \left(J(J+1) - J^{z}(J^{z}-1) \right) - \chi\left(
    (J-\frac{1}{2})(J+\frac{1}{2}) - (J^{z}-\frac{1}{2})(J^{z}-\frac{3}{2})
    \right) \\
    &= \chi \left( J - J^{z} + 1 \right) \leq \chi \left(N_g + 1 \right),
\end{split}
\end{align}
with equality holding iff $J=\tfrac{N_b + N_g}{2}$. We conclude that as atoms dephase with fixed $J^{z}$, $\Delta_\mathrm{SG}$ decreases.

The takeaway from this analysis is that the red dashed curve in Fig.~\hyperref[figS1]{\ref{figS1}b} overestimates $\Delta_\mathrm{SG}$ because it does not take dephasing into account. This is verified by the fact that the computed value of $\Delta_\mathrm{SG}$ (black solid curve) lies below this curve. In the main text, we compare $\Delta_\mathrm{SG}$ against the dissipation-free version of $N_{g}^{\prime\,\mathrm{eff}}$ (black dashed curve) for this reason.
In fact, we see that this curve also overestimates $\Delta_\mathrm{SG}$. This is likely because the initial drive pulse does not place the system in a maximally symmetric state (atoms with different cavity couplings are initialized to a different position on the Bloch sphere).


\putbib
\end{bibunit}

\end{document}